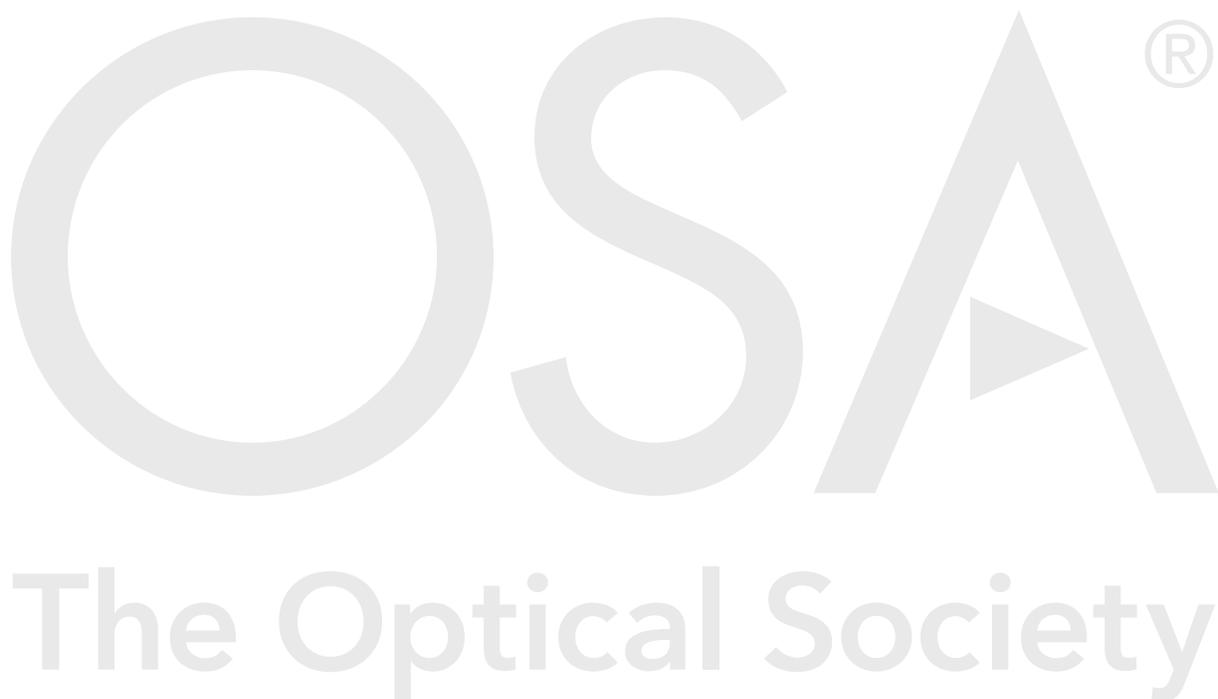

# SOA-MZI photonic sampler with post-distortion linearization technique

**DIMITRIOS KASTRITSIS,**[1,2,*] **THIERRY RAMPONE,**[1] **MAEVA FRANCO,**[1] **KYRIAKOS E. ZOIROS**[2] **AND AMMAR SHARAIHA**[1]

[1]*LabSTICC (UMR CNRS 6285), École Nationale d'Ingénieurs de Brest (ENIB), Brest, France*
[2]*Department of Electrical and Computer Engineering, Lightwave Communications Research Group, Democritus University of Thrace (DUTH), Xanthi, Greece*
*\*dimitrios.kastritsis@enib.fr*

**Abstract:** We present a post-distortion linearization technique for a Semiconductor Optical Amplifier Mach-Zehnder Interferometer (SOA-MZI) photonic sampler. The sampling source is an active mode-locked laser producing 12.6 ps-width pulses with a repetition frequency of 10 GHz. The mathematical model for the linearization technique is presented, and then evaluated for the quasi-static regime, i.e. sampling continuous-wave signals, and for the dynamic regime, i.e. sampling sinusoidal signals. A significant improvement in terms of Total Harmonic Distortion (THD) equal to 23.4 dB is observed for a modulation index equal to 80% in the quasi-static regime, matching the highest observed THD improvement in the dynamic regime.

## 1. Introduction

Analog Radio over Fiber (ARoF) links provide a set of performance advantages for handling microwave signals, and in particular the large bandwidth, the low attenuation, the high immunity to electromagnetic interference and the decreased size and weight of the fiber transmission medium. As a result, ARoF photonic links find application in 5G, radar, Internet of Things (IoT), electronic warfare, and analog-to-digital conversion [1–8]. Additionally, optical links can be designed to realize signal processing functions in order to increase the flexibility and boost the performance of such systems.

Sampling is a fundamental signal processing function of ARoF. Initially, the sampling process was realized in the electrical domain, which impeded the exploitation of its full potential due to the limited electronic bandwidth and the timing jitter of electronic oscillators [7, 8]. Lately, however, sampling in the optical domain has become an attractive technological alternative owing to its ability to combine the advantages of ARoF links with the possibility of using a low jitter mode-locked laser [9, 10] as the sampling signal [8, 11, 12]. Moreover, by being part of an ARoF scheme, optical sampling allows for the optical source of the signal to be sampled and of the sampling signal to be physically distant from each other, thus enhancing the versatility of the process.

The difference in performance potential between electrical and optical samplers can be highlighted by comparing some key performance metrics of the two modules. More specifically, an electrical sampler employed in a state of the art 6 bit Analog to Digital Converter (ADC) of flash architecture has a sampling frequency of 10.3 GS/s with a bandwidth ranging from 3 to 6 GHz [13]. On the contrary, a Mach-Zehnder Modulator (MZM)-based optical sampler in an implementation of time-stretch ADC has a 480 GS/s sampling rate with a bandwidth of 50 GHz [14]. More complex electronic ADC architectures employing 64 elementary ADCs have been proposed in [15] achieving a 56 Gs/s sampling rate with 31.5 GHz analog bandwidth but at the expense of a very high complexity.

Nevertheless, because the sampler characteristic function is not linear, the distortion products that appear at the output degrade the performance of the link, especially for signals with high modulation indices [12, 16, 17]. Thus, mitigation of nonlinear distortion i.e., linearization, is the key in order to improve the performance and expand the exploitable input power range of the photonic sampler [12, 16].

A significant performance improvement of the optical sampling process has been achieved by employing a post-distortion linearization technique and applying it to an MZM [12, 16]. The potential of this technique gave rise to the idea of applying a similar post-distortion linearization method for the first time to a Semiconductor Optical Amplifier Mach-Zehnder Interferometer (SOA-MZI).

Photonic sampling employing a SOA-MZI has been used for the application of photonic microwave mixing in [17–23], since the SOA-MZI is an all-optical module which uniquely combines a low excitation power, a high extinction ratio and a high switching efficiency with the simultaneous signal amplification [24, 25].

Total Harmonic Distortion (THD) is a fundamental metric for almost any communication system, especially in the context of broadband RF systems [26, 27]. High THD may signify interference for other equipment or require elaborate filtering at the expense of increased cost and complexity of the ARoF link. Recently, THD has been evaluated for the SOA-MZI photonic sampler in [17, 28]. Applying the post-distortion linearization technique to this case aims exactly at minimizing THD of the SOA-MZI photonic sampler.

In this paper, we present, for the first time to our knowledge, the proof-of-concept of a post-distortion linearization method for a SOA-MZI sampler, and we demonstrate through theory and experiment its feasibility, potential and effectiveness. The remainder of the paper is organized as follows. In Section 2, we describe the principle of operation and formulate the mathematical model of the SOA-MZI post-distortion linearization method. In Section 3, we provide the experimental setups, define the conditions used for the validation of the method and introduce empirical correction coefficients for obtaining the final form of the power estimation equation. In Section 4, we present optical sampling results in the time domain, explain how the post-distortion method is applied and compare the THD, before and after the application of the post-distortion method, for sinusoidal signals of different frequencies and modulation indices. Additionally, a THD estimation is given, assuming that ideal sinusoidal signals at the SOA-MZI input are sampled, so as to establish the potential of the method in terms of maximum THD improvement.

## 2. Principle of operation of the post-distortion linearization method

Fig. 1 illustrates the principle of operation of the post-distortion linearization method in a SOA-MZI. In this method, the function of sampling is obtained using a modulation architecture presented for the first time in [17] and in detail in [23]. In the SOA-MZI, a sampling pulse train of instantaneous power $P_C$, center wavelength $\lambda_C$, repetition rate $f_C$ and equal amplitude pulses generated by a mode-locked laser, is applied at input C, while a sinusoidal signal to be sampled of instantaneous power $P_A$, modulation index $MI$, center wavelength $\lambda_A$ and frequency $f_A$ is applied at input A. These signals interact inside the SOA-MZI via Cross-Phase Modulation (XPM), so that the pulse train is amplitude modulated in accordance with the variations of the amplitude of the signal to be sampled. The result of this process at $\lambda_C$ is obtained at the SOA-MZI outputs I and J, which are complementary, i.e. the outcomes therein exhibit opposite variations.

The use of the Modulation architecture [23] means that the sampling signal with a repetition frequency $f_C$ is only amplified, so $f_C$ can be very high due to the high optical amplification bandwidth of the SOA, despite the dynamics of the SOA carriers. On the other hand, the signal to be sampled, at the frequency $f_A$, must be below the cutoff frequency of the XPM process, which depends on the dynamics of the SOA carriers. As a consequence, the analog bandwidth of the SOA-MZI photonic sampler is limited by the XPM bandwidth, which is between 5 and 6 GHz for the specific operating conditions used in this paper [28].

In our photonic sampling architecture, we use a technique referred to as post-distortion linearization that combines at each sampling instant the power for the complementary ports I and J to invert the SOA-MZI transfer function. In order to achieve this, the two signals emerging from ports I and J are detected by an oscilloscope and the recorded datasets are stored. Then, a signal processing unit exploits the mathematical model that is formulated in subsection 2.1 and subsequently empirically corrected by a simple optimization process based on experimental results in subsection 3.2. This is done in order to linearize the output and estimate the power of the signal to be sampled, $P_A$, using the stored datasets from ports I and J, the operating parameters of the SOA-MZI as well as the empirical correction coefficients.

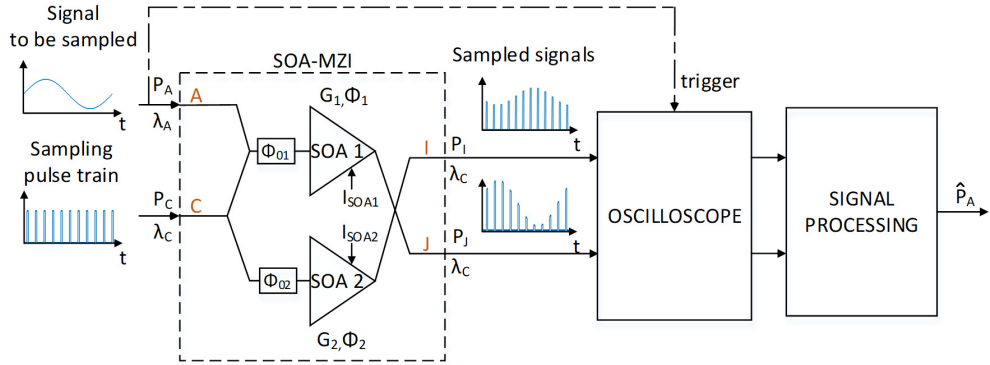

Fig. 1. Conceptual diagram for application of post-distortion linearization method in a SOA-MZI photonic sampler.

## 2.1 Mathematical model

The SOA-MZI input power injected at port C is assumed to comprise a train of Gaussian-shaped pulses, and hence it can be expressed in the time domain as [29]:

$$P_C = P_{C,peak} \sum_{\mu} \exp\left(-4\ln 2 \left(\frac{t - \frac{\mu}{f_C}}{T_{fwhm}}\right)^2\right), \quad (1)$$

where $P_{c,peak}$ is the peak power, $f_C$ is the repetition frequency or sampling rate and $T_{fwhm}$ is the full-width at half-maximum pulse width of each $\mu$-th distinct pulse. The sinusoidal input power injected at port A can be expressed in the time domain as [23]:

$$P_A = P_{A,avg}\left[1 + m\cos(2\pi f_A t)\right], \quad (2)$$

where $P_{A,avg}$ is the average power of $P_A$ and $m$ is the modulation index.

The SOA-MZI output power emerging from ports I and J at the wavelength $\lambda_C$ is given by the following equations [23, 30]:

$$P_I = \frac{1}{8}\left[G_1 + G_2 + 2\sqrt{G_1 G_2}\cos(\Phi_1 - \Phi_2 + \Phi_0)\right]P_C, \qquad (3)$$

$$P_J = \frac{1}{8}\left[G_1 + G_2 - 2\sqrt{G_1 G_2}\cos(\Phi_1 - \Phi_2 + \Phi_0)\right]P_C, \qquad (4)$$

where $G_1$, $G_2$ are the SOA1 and SOA2 optical gains, respectively, at $\lambda_C$, $\Phi_1$-$\Phi_2$ is the phase difference induced by XPM between the two arms of the interferometer, and $\Phi_0$ is the static phase shift equal to $\Phi_{01}$-$\Phi_{02}$, which is considered to be zero in our operating conditions. The analysis that follows is based on the assumption that the physical parameters of the two SOAs incorporated in the MZI are identical, as well as that $I_{SOA1} = I_{SOA2} = I_{SOA}$. SOAx gain $G_x$ is defined as follows:

$$G_x = \exp(g_{n,x} L), \quad x \in \{1,2\}, \qquad (5)$$

where $L$ is the active region length and $g_{n,x}$ is the SOAx net gain which is defined by:

$$g_{n,x} = \Gamma g_{m,x} - \alpha_{int}, \qquad (6)$$

where $\Gamma$ is the active zone optical confinement factor, $\alpha_{int}$ is the internal loss and $g_{m,x}$ is the SOAx modal gain defined by:

$$g_{m,x} = \alpha_C(N_x - N_0), \qquad (7)$$

where $\alpha_C$ is the peak-gain coefficient, $N_x$ is the SOAx carrier density and $N_0$ is the carrier density at transparency. Using Eq.(3), Eq.(4) and $\Phi_0 = 0$, the difference between $P_I$ and $P_J$ is:

$$P_I - P_J = \frac{1}{2}\sqrt{G_1 G_2}\cos(\Phi_1 - \Phi_2)P_C. \qquad (8)$$

Now we can express the phase difference between the two arms of the SOA-MZI in relation to the output powers, $P_I$ and $P_J$, and the input power of the sampling pulse train, $P_C$:

$$\Phi_1 - \Phi_2 = \cos^{-1}\left(\frac{2}{\sqrt{G_1 G_2}} \frac{P_I - P_J}{P_C}\right). \qquad (9)$$

Taking the sum of the powers $P_I$ and $P_J$ and expressing the input power $P_C$ in relation to the output powers $P_I$, $P_J$ and the gains $G_1$ and $G_2$ results in:

$$P_C = 4\frac{P_I + P_J}{G_1 + G_2}. \qquad (10)$$

Using Eq.(10) and solving for $G_1$ we get the following expression:

$$G_1 = 4\frac{P_I + P_J}{P_C} - G_2. \tag{11}$$

Substituting the expression for $P_C$ from Eq.(10) in Eq.(9) we have:

$$\Delta\Phi = \Phi_1 - \Phi_2 = \cos^{-1}\left[\frac{1}{2}\frac{P_I - P_J}{P_I + P_J}\frac{G_1 + G_2}{\sqrt{G_1 G_2}}\right]. \tag{12}$$

In order to couple the phase difference $\Delta\Phi$ with the input signals injected into ports A and C at wavelengths $\lambda_A$ and $\lambda_C$ and of power $P_A$ and $P_C$, respectively, we use the carrier density rate for SOA1 and for SOA2. Carrier density rate for SOA1 is given by [31]:

$$\frac{dN_1}{dt} = \frac{I_{SOA}}{qV} - R_A(N_1) - R_C(N_1) - R_{nr}(N_1) - R_{ASE}(N_1), \tag{13}$$

where $I_{SOA}$ is the SOA1 bias current, q is the elementary electrical charge, $V$ is the active zone volume of SOA1, $R_A(N_1)$ and $R_C(N_1)$ are the recombination rates due to the amplification of the optical carriers at wavelengths $\lambda_A$ and $\lambda_C$, respectively. $R_{nr}(N_1)$ is the non-radiative recombination rate and $R_{ASE}(N_1)$ is the recombination rate due to the Amplified Spontaneous Emission (ASE). The cumulative influence of $R_{nr}$ and $R_{ASE}$ can be taken into account by introducing the term $R$:

$$R(N_1) = R_{nr}(N_1) + R_{ASE}(N_1) = \frac{N_1}{\tau}, \tag{14}$$

where $\tau$ is the SOA effective carrier lifetime derived from $\tau_{nr}$, which is related to $R_{nr}$ and $\tau_{ASE}$, which is related to $R_{ASE}$, employing the equation [32]:

$$\frac{1}{\tau} = \frac{1}{\tau_{nr}} + \frac{1}{\tau_{ASE}}. \tag{15}$$

Using the approximation that the SOA gain at wavelength $\lambda_A$ is the same as that at $\lambda_C$, as in [23], $R_A(N_1)$ is given by:

$$R_A = \frac{1}{2}\frac{\lambda_A}{hc}\frac{\Gamma g_{m,1}}{g_{n,1}}\frac{G_1 - 1}{V}P_A, \tag{16}$$

where $h$ is Plank's constant and $c$ is the speed of light in vacuum. Similarly, $R_C(N_1)$ is given by:

$$R_C = \frac{1}{4}\frac{\lambda_C}{hc}\frac{\Gamma g_{m,1}}{g_{n,1}}\frac{G_1 - 1}{V}P_C. \tag{17}$$

Applying Eq.(14), Eq.(16) and Eq.(17) to Eq.(13) and using the approximation $\Gamma g_{m,1} = g_{n,1}$, as in [23], we obtain:

$$\frac{dN_1}{dt} = \frac{I_{SOA}}{qV} - \frac{N_1}{\tau} - \frac{1}{2}\frac{\lambda_A}{hcV}P_A(G_1-1) - \frac{1}{4}\frac{\lambda_C}{hcV}P_C(G_1-1). \tag{18}$$

We develop our model on the assumption of a static SOA regime, which is defined by $\frac{dN_1}{dt} = 0$. Although by making such an approximation it is expected that the SOA and SOA- MZI response will deviate from the real one, still the inevitable discrepancy can be kept acceptable within the considered operating range of use of this device and module, respectively. Furthermore, the model's accuracy can be improved by applying an empirical correction approach, as it will be described in Section 3. In this manner, we manage to formulate a sufficiently precise model while preserving its simplicity and its ability to produce analytic solutions.

We can express the SOA1 carrier density in the static regime by the following equation:

$$N_1 = \frac{1}{\Gamma\alpha_C L P_{sat}}\left[\frac{hcI_{SOA}}{\lambda_C q} - \frac{1}{4}P_C(G_1-1) - \frac{1}{2}\frac{\lambda_A}{\lambda_C}P_A(G_1-1)\right], \tag{19}$$

where $P_{sat}$ is defined by [23]:

$$P_{sat} = \frac{hcV}{\lambda_C \Gamma\alpha_C L\tau}. \tag{20}$$

For SOA2, provided that its physical parameters ($\Gamma$, $\alpha_c$, $L$, $P_{sat}$) are the same as for SOA1, and given the fact that a single optical signal passes through it at $\lambda_C$, we have, after following the same analysis as for SOA1:

$$N_2 = \frac{1}{\Gamma\alpha_C L P_{sat}}\left[\frac{hcI_{SOA}}{\lambda_C q} - \frac{1}{4}P_C(G_2-1)\right]. \tag{21}$$

From Eq.(19) and Eq.(21), we obtain the difference between the SOA1 and SOA2 carrier density by:

$$N_1 - N_2 = \frac{1}{\Gamma\alpha_C L P_{sat}}\left[\frac{1}{4}P_C(G_2-G_1) - \frac{1}{2}\frac{\lambda_A}{\lambda_C}P_A(G_1-1)\right]. \tag{22}$$

The linearized expression of the SOA active zone effective refractive index as a function of carrier density is given by [33]:

$$n_{e,x} = n_{e0} + \frac{\partial n_e}{\partial N}(N_x - N_0), \; x \in \{1,2\}, \tag{23}$$

where $n_{e0}$ is the refractive index at transparency and $N_0$ is the SOA carrier density at transparency. Using Eq.(23) and Eq.(22), we can calculate the phase difference between the SOAs' outputs. This difference at wavelength $\lambda_c$ is given by:

$$\Phi_1 - \Phi_2 = \frac{\alpha_H}{2P_{sat}}\left[\frac{1}{4}P_C(G_1-G_2) + \frac{1}{2}\frac{\lambda_A}{\lambda_C}P_A(G_1-1)\right], \quad (24)$$

where $\alpha_H$ is the linewidth enhancement, or Henry's, factor, defined by:

$$\alpha_H = -\frac{4\pi}{\lambda_C}\frac{\partial n_e/\partial N}{\partial \Gamma g_{m,x}/\partial N} = -\frac{4\pi}{\lambda_C \Gamma \alpha_C}\frac{\partial n_e}{\partial N}. \quad (25)$$

Although Eq.(12) has been derived using time domain equations, i.e., $P_I$, $P_J$, are the instantaneous powers at ports A, I and J and $G_1$, $G_2$ are the instantaneous gains of SOA1 and SOA2, respectively, Eq.(24) has been derived using their average values. Equating Eq.(12) and Eq.(24) permits us to express $P_A$ as a function of measured parameters (powers and gains) and thus approximate it empirically by:

$$P_A = \frac{\lambda_C}{\lambda_A}\frac{2}{G_1-1}\left\{\frac{2P_{sat}}{\alpha_H}\cos^{-1}\left[\frac{1}{2}\frac{P_I-P_J}{P_I+P_J}\frac{G_1+G_2}{\sqrt{G_1G_2}}\right]-\frac{1}{4}P_C(G_1-G_2)\right\}. \quad (26)$$

It should be underlined that Eq.(26) is universal and thus it applies in general for a SOA-MZI. This holds since the specific expression is a function of fundamental parameters, i.e., input signal parameters such as the power $P_C$ and the wavelengths $\lambda_A$ and $\lambda_C$, SOA physical parameters, such as $P_{sat}$ and $a_H$, SOA operating parameters, such as $G_1$, $G_2$, and SOA-MZI output powers, $P_I$, $P_J$.

## 3. Linearization of sampled continuous-wave signals

### 3.1 Experimental setup

Fig. 2 shows the experimental setup for conducting a quasi-static characterization of the SOA-MZI (CIP 40G-2R2-ORP) response in order to choose the average power of $P_A$. This type of characterization incorporates the dynamics and the behavioral changes caused to the SOA-MZI response by the utilization of optical pulses, as opposed to a pure static case. An Optical Pulse Clock (OPC) source, which is an active mode-locked laser (Pritel model UOC-E-05-20), is driven by an RF generator at a frequency $f_C$ equal to 10 GHz and provides an optical pulse train of 12.6 ps full-width at half-maximum pulse width. The signal produced by the OPC, centered at $\lambda_C$ = 1557.4 nm and having its average power adjusted by an optical attenuator to be constant at – 15 dBm, serves as the sampling signal that enters SOA-MZI port C. The two phase-shifters $\Phi_{01}$ and $\Phi_{02}$, integrated to the SOA-MZI package, as well as the polarization controller (PC1) are tuned in such a way that the optical power at output port J is minimum when the power at input port A is zero. A continuous-wave signal tuned at $\lambda_A$ = 1550 nm enters SOA-MZI port A and serves as the signal to be sampled. The average power of the pulses at outputs I and J was measured, after optical filtering centered at $\lambda_C$ with a 3 dB optical bandwidth equal to 0.7 nm, using an optical power meter.

Fig. 3 shows the response of the SOA-MZI at outputs I and J as a function of the average value of $P_A$. We can observe the nonlinearity of the SOA-MZI response, which hence underlines the need to improve the linearity of the response as a function of the power $P_A$ of the signal to be sampled. The maximum range of injected $P_A$ values in which the linearization method can be exploited lies between a response minimum and maximum, due to the fact that the argument of the $\cos^{-1}$ function in Eq.(26) should not exceed ±1. We choose the first such range between – 30 Bm and – 5 dBm. The obvious choice for the average value of $P_A$ is at the middle of this range (i.e. – 8 dBm), so that an intensity-modulated signal with a *MI* of 100% can take power values within the whole permissible range of $P_A$ values. To ensure that we never surpass, even instantaneously, the maximum in case of an eventual fluctuation due to temperature changes, the average value of $P_A$ is chosen 1 dB lower than the middle of the specified range, i.e. – 9 dBm.

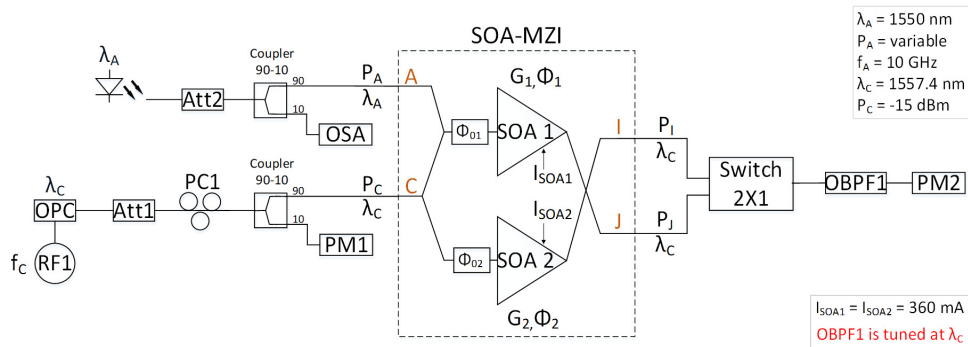

Fig. 2. Quasi-static characterization experimental setup in order to choose the average power of $P_A$. Att: Attenuator. OSA: Optical Spectrum Analyzer. PC: Polarization Controller. RF: Radio Frequency generator. PM: Power Meter. OPC: Optical Pulse Clock source. OBPF: Optical Band-Pass Filter.

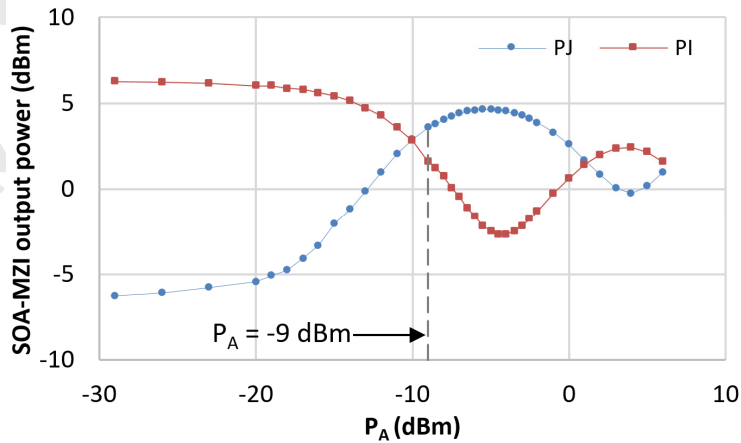

Fig. 3. Quasi-static response of SOA-MZI.

*3.2 Linearization and model corrections*

The linearization method is applied on the sampling of continuous-wave signals shown in Fig. 3. First, $G_2$ is measured optically and found equal to 26.8 dB. Afterwards, the average values of $P_I$, $P_J$ and $P_C$ as well as the previously measured $G_2$ are used in order to recover the values of $G_1$ from Eq.(11). With $G_1$ at every point in Fig. 3 known, the nonlinear phase shift $\Delta\Phi$ can also be recovered from Eq.(12). Finally, the estimated $\hat{P}_A$ is given by Eq.(26). The value of the $P_{sat}/a_H$ ratio is 5.82 mW. It is worth noting that the optical gain $G_2$, at the lower arm of the SOA-MZI, does not depend on changes in $P_A$, therefore $G_2$ only needs to be measured once at the beginning of the process.

Observing the results in Fig. 4(a), we notice that we have an offset error and a slope gain error. These differences are due to the simplifications made throughout the above theoretical development, such as the perfect symmetry between the SOA-MZI arms, the identical physical parameters $\Gamma$, $\alpha_c$, $L$ and $P_{sat}$ for SOA1 and SOA2, the negligible SOAs internal losses $a_{int}$ compared to $\Gamma g_{m,1}$, i.e. $\Gamma g_{m,1} = g_{n,1}$, and $\Phi_0 = 0$. Thus to correct the slope and offset of the estimated input power, we apply two empirical correction coefficients denoted by '$gco$' and '$off$', respectively, which are estimated using the experimental results. To this end, $P_A$ can be estimated by the use of the following expression:

$$\hat{P}_A = gco\left[\frac{\lambda_C}{\lambda_A}\frac{2}{G_1-1}\left(\frac{2P_{sat,C}}{\alpha_H}\Delta\Phi - \frac{1}{4}P_C(G_1-G_2)\right)\right] + off. \qquad (27)$$

In Fig. 4(b), we introduce '$gco$' and '$off$' coefficients as in Eq.(27), and then we optimize their values by using a least squares approximation in order to minimize the error between $\hat{P}_A$ and $P_A$. The optimized values are equal to 0.48 for '$gco$' and equal to – 17.9 μW for '$off$'.

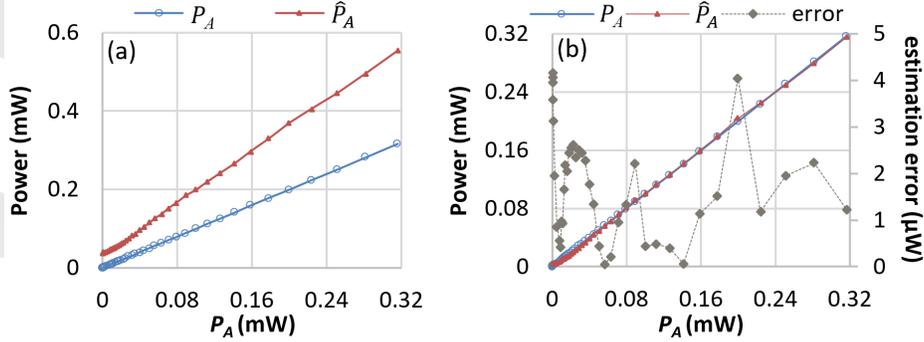

Fig. 4. Comparison of the estimated $\hat{P}_A$ after the application of the post-distortion linearization method and the injected $P_A$ for the quasi-static results in Fig. 3 without the use of coefficients (a) and with '$gco$' = 0.48 and '$off$' = – 17.9 μW (b).

Utilizing the linearization technique for continuous-wave signals, while employing the '$gco$' and '$off$' empirical correction coefficients as in Fig. 4(b), we obtain an average error as low as between $P_A$ and $\hat{P}_A$ for the whole range between 0.0001 mW and 0.32 mW or equivalently between – 40 dBm and – 5 dBm.

## 4. Linearization of sampled sinusoidal signals

### 4.1 Experimental setup

The linearization measurement setup in the dynamic regime is shown in Fig. 5. The SOA-MZI sampling pulse signal is produced, and $\Phi_{01}$, $\Phi_{02}$ and PC1 are adjusted, in the same way as in quasi-static regime measurement. A continuous-wave signal tuned at $\lambda_A$ = 1550 nm is intensity-modulated by an MZM, which is biased at the quadrature point with $V_{bias}$ = 3.1 V and produces a sinusoidal Double Sideband Full Carrier (DSB-FC) signal whose frequency $f_A$ is provided by an RF generator. The DSB-FC signal serves as the signal to be sampled that enters SOA-MZI port A. The average power of $P_A$ is chosen at – 9 dBm, as previously mentioned.

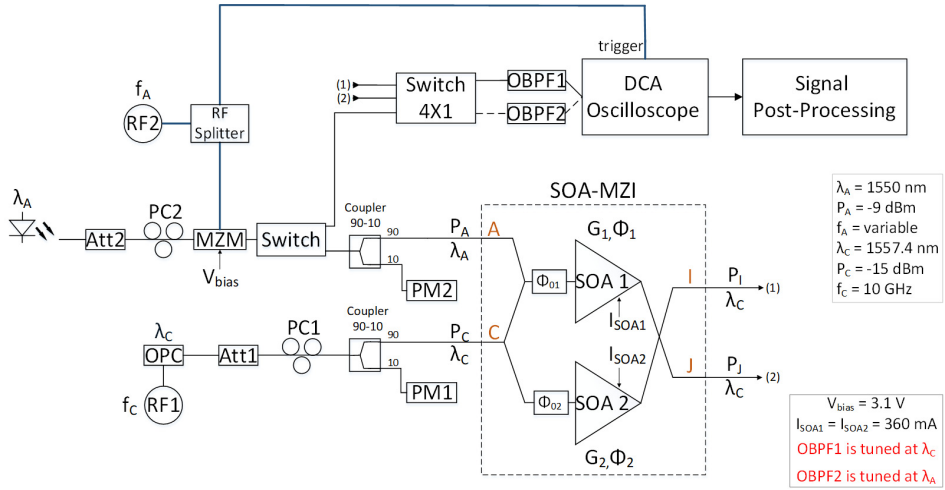

Fig. 5. Linearization measurement setup. MZM: Mach-Zehnder Modulator. DCA: Keysight N1000A Digital Communication Analyzer.

The power of the pulses at outputs I and J as well as the power at input A are measured after optical filtering using a sampling oscilloscope (Keysight N1000A DCA-X with N1030A calibrated optical plugin), and the datasets of the three signals are recorded and saved. Additionally, $G_2$ is measured optically and found equal to 26 dB. Also the value of $P_C$ corresponds to the pulses peak power. An example of a signal to be sampled with $f_C$ = 1 GHz and $MI$ = 60% is shown in Fig. 6(a), and of the sampling signal in Fig. 6(b). The corresponding sampled signals at port I and J are shown in Fig. 7(a) and Fig. 7(b), respectively.

All oscilloscope measurements have been obtained using averaging acquisition mode of 1024 values. The optical plugin RMS noise, which is the main source of error in the oscilloscope, is 45 µW. Due to averaging the measured RMS error of the time domain results is reduced to $45/\sqrt{1024}$ = 1.41 µW.

### 4.2 Sampled sinusoidal signals and linearization

The first step of the signal's treatment is the isolation of pulses' peaks for port I and port J depicted by the dashed line curves in Fig. 7(a) and Fig. 7(b), respectively. The signal processing that follows is performed on the pulse peaks which are the samples that result from the sinusoidal signals' sampling. The peak power of $P_C$ input pulses, $P_{C,peak}$, is approximately calculated from:

$$P_{C,peak} = \frac{P_{C,ave}}{f_C T_{fwhm}}, \tag{28}$$

where $P_{C,ave}$ is the average power of $P_C$, $f_C$ is the sampling rate and $T_{fwhm}$ is the full-width at half-maximum pulse width. In our case, $P_{C,ave}$ = - 15 dBm, $f_C$ = 10 GHz and $T_{fwhm}$ = 12.6 ps, and thus $P_{C,peak}$ is approximately 0.25 mW or -6 dBm.

The next step is the recovery of dynamic gain $G_1$ using Eq.(11) and of the dynamic nonlinear phase shift $\Delta\Phi$ from Eq.(12), corresponding to every pulse peak or sample. Finally, $\hat{P}_A$ is calculated using Eq.(27), while the linearization model parameters and '*gco*' and '*off*' empirical correction coefficients are optimized to minimize the mean squared error between $P_A$ and $\hat{P}_A$. A visual comparison between $P_A$ and $\hat{P}_A$ for $f_C$ = 1 GHz with $MI$ = 40%, for $f_C$ = 1 GHz with $MI$ = 60%, for $f_C$ = 0.25 GHz with $MI$ = 80% and for $f_C$ = 0.5 GHz with $MI$ = 80%, is shown in Fig. 8(a), (b), (c) and (d), respectively. The average percentage error in Fig. 8(a) is 60% before the application of the linearization method, whereas it is 8.1% after the application. Similarly, the average percentage error in Fig. 8(b) is 15.3% before the application of the linearization method, whereas it is 13.4% after the application. The corresponding values before and after the application of the linearization method are 52.3% and 8% in Fig. 8(c), and 66.6% and 13% in Fig. 8(d).

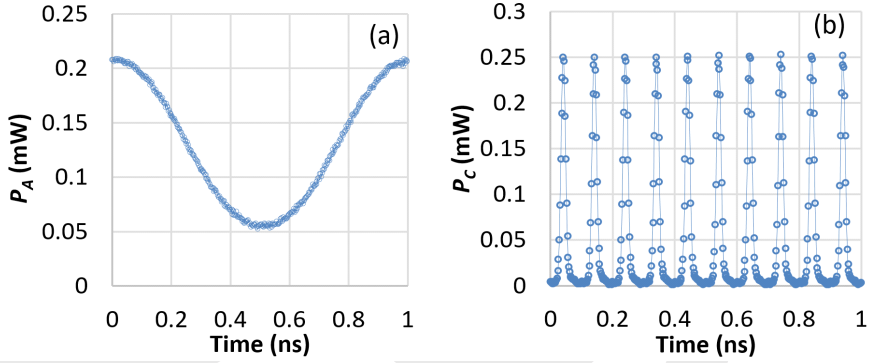

Fig. 6. Input power at SOA-MZI port A (a) and port C (b).

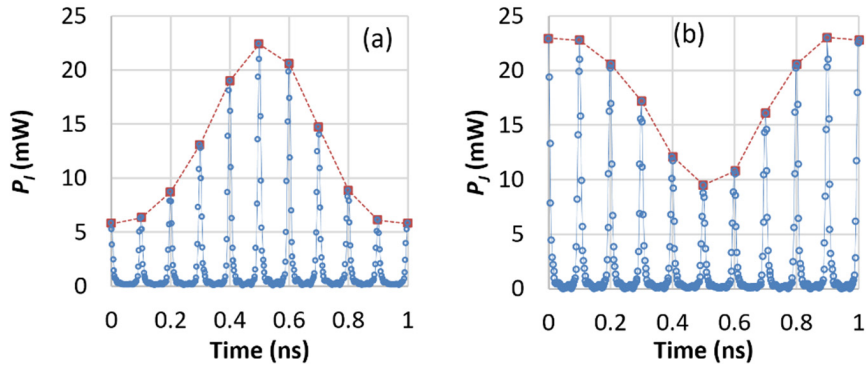

Fig. 7. Power at port I (a) and port J (b) of the sampled signal for $f_A$ = 1 GHz and $MI$ = 60%. The first step of the signal's treatment is the isolation of pulses' peaks for port I illustrated by the dashed line in (a) and for port J illustrated by the dashed line in (b). All oscilloscope measurements have been obtained using averaging acquisition mode of 1024 values.

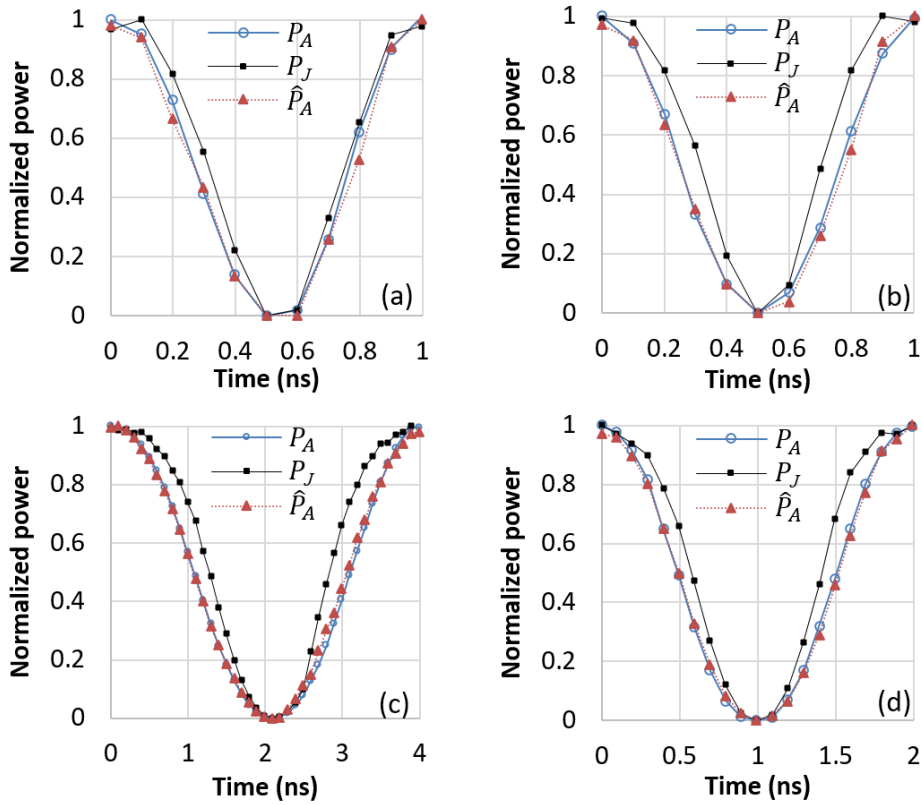

Fig. 8. Comparison of the normalized values between the estimated $\hat{P}_A$, the injected $P_A$ and the direct SOA-MZI output $P_J$. (a) $f_A = 1$ GHz and $MI = 40\%$. (b) $f_A = 1$ GHz and $MI = 60\%$. (c) $f_A = 0.25$ GHz and $MI = 80\%$. (d) $f_A = 0.5$ GHz and $MI = 80\%$.

## 5. THD linearization performance

In order to obtain the electrical spectrum of $P_J$, $\hat{P}_A$ and $P_A$, a Fast Fourier Transform (FFT) is performed on the respective time domain signals. An example of the electrical spectrum of $P_J$, $\hat{P}_A$ and $P_A$ for the case of $f_C = 1$ GHz and $MI = 60\%$ is shown in Fig. 9(a), (b) and (c), respectively. We observe a clear improvement in terms of harmonic distortion comparing the electrical spectra after (Fig. 9(b)) and before (Fig. 9(a)) the application of the post-distortion method. The improvement is mostly attributed to the reduction of Harmonic Distortion 2 (HD2) at 2 GHz. The linearity of the photonic sampling mixer is assessed using THD as performance metric. For the results presented in the following, the THD is defined for the cases of $f_A$ equal to 0.25 GHz, 0.5 GHz and 1 GHz as the sum of harmonic distortion frequency components of order 2 to 4 (HD2-4) divided by the fundamental target frequency component. However, for the case of $f_A$ equal to 2 GHz the THD is defined as the HD2, which is the most significant distortion product, divided by the fundamental target frequency component. This discrepancy is due to the sampling frequency $f_C$, which is 10 GHz, meaning that as we consider exclusively the pulses' peaks and for a Nyquist sampling rate equal to 10 GHz, the highest frequency provided by FFT is 5 GHz.

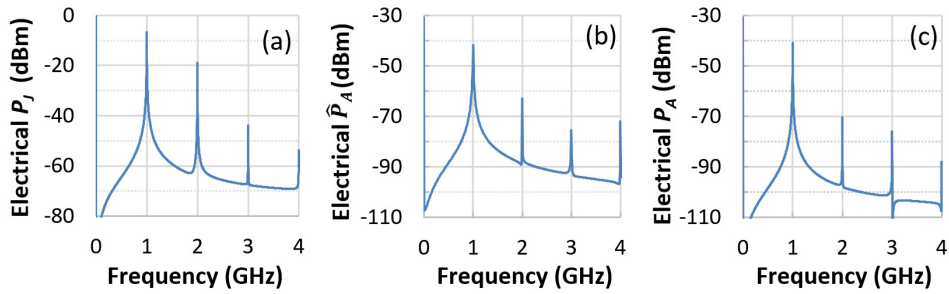

Fig. 9. Electrical spectrum of $P_J$(a), $\hat{P}_A$(b) and $P_A$ (c) calculated with Fast Fourier Transform (FFT) for $f_A$ = 1 GHz and $MI$ = 60 %.

*5.1 THD results as a function of the modulation index of the signal to be sampled*

Fig. 10(a) shows a comparison between the THD of the signal after the application of the linearization method, denoted as 'THD post-pr' to distinguish it from that at the SOA-MZI direct output J, denoted as 'THD PJ', and to that of the signal to be sampled at the input of the SOA-MZI, denoted as 'THD modulator', as a function of the modulation index $MI$ of the signal to be sampled. The carrier frequency $f_A$ is equal to 250 MHz. We observe that the THD improvement achieved by the linearization technique is moderate for a $MI$ equal to 20%, significant for a $MI$ equal to 40% and to 60%, and extremely high and equal to 23.9 dB for a $MI$ equal to 80%.

Fig. 10(b) shows the same comparison as in Fig. 10(a) but for $f_A$ equal to 1 GHz. In this figure, we notice that the maximum THD improvement provided by the method is significant and equal to 9.1 dB for $MI$ equal to 60%. It should be noted that the error bars in the THD lines correspond to the variation in the THD values that could be caused by small random fluctuations of the order of 5% in the $P_J$, $\hat{P}_A$ and $P_A$ time domain results. This error is equivalent to ±1.5 dB for the THD post-pr, THD PJ and THD modulator lines and equivalent to ±3 dB for the THD improvement line. Moreover, it is worth noting that from a theoretical point of view the THD results at the complementary output I are expected to be similar to that at direct output J. Therefore, we have made the choice to compare the linearized results to the results at port J.

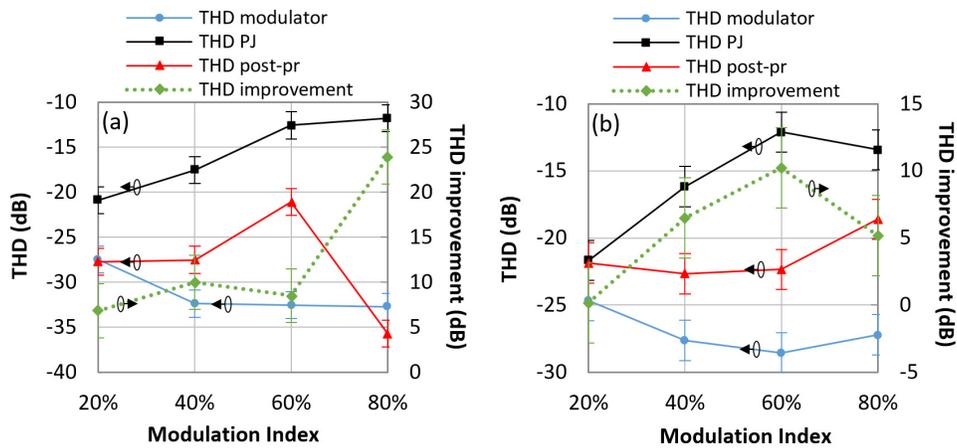

Fig. 10. Comparison between THD of linearized and SOA-MZI's direct output as a function of the modulation index $MI$ of the signal to be sampled. (a) $f_A$ = 250 MHz. (b). $f_A$ = 1 GHz.

In order to establish the potential of the method, an estimation of the THD can be made, based on the quasi-static results in Fig. 3 and in Fig. 4(b). $P_J$ and $\hat{P}_A$ responses are curve-fitted through interpolation to a 12$^{th}$ order polynomial. The THD is estimated assuming an ideal sinusoidal signal at port A with an average optical power equal to – 9 dBm. A FFT is applied on the time domain estimations of the two signals in order to produce their electrical spectra and subsequently calculate their THD and the THD improvement provided by the technique as shown in Fig. 11. It should be noted that the THD estimation of the quasi-static regime does not take into account the dynamic phenomena that take place into the SOA-MZI, and this estimation is valid only for low frequencies, while it is compromised at high frequencies. Also, it should be emphasized that the quasi-static regime results are based on measurements using an optical power meter, whereas the dynamic regime results are based on measurements using a wide bandwidth oscilloscope.

We observe in Fig. 11 that the THD improvement is almost constant around 20 dB from *MI* equal to 20% to *MI* equal to 60%, and also we notice that a peak of 23.8 dB is observed for *MI* equal to 76%.

Comparing the THD response for $f_A$ equal to 250 MHz in Fig. 10(a) with the THD quasi-static response in Fig. 11, there is a matching between the THD post-pr, THD P$_J$ and THD improvement values for *MI* equal to 80% when we take into account the 5% error bars' lower limit. However, for *MI* equal to 20%, 40% and 60% the THD post-pr, THD $P_J$ and THD improvement values in Fig. 11 compared to those in Fig. 10(a) are significantly different even if we take into account the error bars. A possible explanation for this difference is that as the signal to be sampled modulation index *MI* decreases from 80%, the RMS error of the oscilloscope becomes more significant compared to the variation of $P_J$, $P_I$ and $P_A$ signals, thus inducing an additional error in the linearization process.

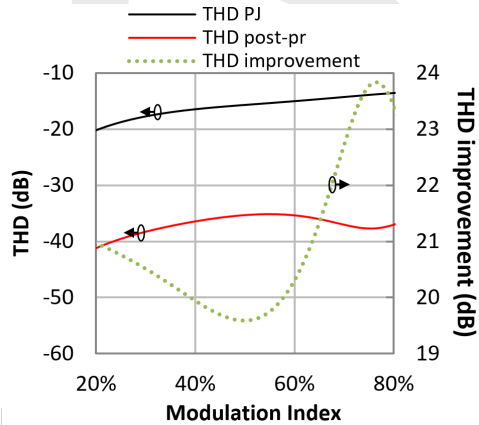

Fig. 11. Comparison between estimated THD of linearized and SOA-MZI's direct output J as a function of the modulation index *MI* of the signal to be sampled. The THD estimation is based on the quasi-static results.

*5.2 THD results as a function of the frequency of the signal to be sampled*

Fig. 12(a), (b) and (c) show the THD of the signal after the application of the linearization method, in comparison to that at the SOA-MZI direct output J and the THD of the signal to be sampled, as a function of the frequency $f_A$ of the signal to be sampled, for different modulation indices 40%, 60% and 80%, respectively. As a general rule, the effectiveness of the method decreases as the frequency increases due to the fact that the model's equations presented in

Section 2 are based on the assumption that $\frac{dN_1}{dt} = 0$. In other words, the linearization equations are best adapted for a static or a quasi-static regime, whereas in a dynamic regime there is a growing error as the frequency increases. However, this error is not prohibiting for the application of the linearization equations on dynamic signals and, as a matter of fact, a significant THD improvement of at least 5.5 dB, taking into account the error bars' lower limit, is observed for $f_A$ = 250 MHz and for $f_A$ = 500 MHz with $MI$ = 40%, 60% and 80%, as well as for $f_A$ = 1 GHz with $MI$ = 60%. For all the other cases, the benefit of using the linearization technique is always tangible but not so pronounced.

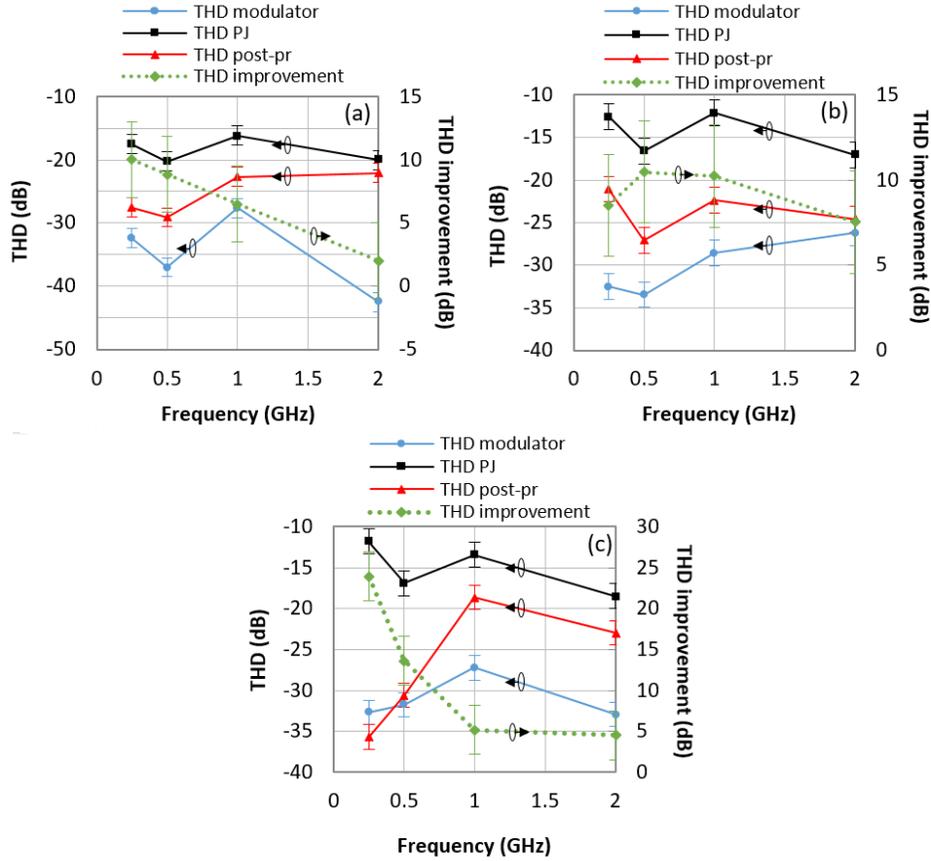

Fig. 12. Comparison between THD of linearized and SOA-MZI's direct output as a function of the frequency $f_A$ of the signal to be sampled. (a) $MI$ = 40%. (b) $MI$ = 60%. (c) $MI$ = 80%.

## 6. Conclusion

In conclusion, we have presented, for the first time to our knowledge, a post-distortion linearization method and demonstrated through theory and experiment its feasibility, potential and effectiveness for a SOA-MZI photonic sampler. Through application of the method to a quasi-static measurement, the estimated potential THD improvement was found to be higher than 19.6 dB for sinusoidal signals with $MI$ between 20% and 80%. This high potential THD improvement is verified through sampling and application of the method to a sinusoidal signal at 250 MHz with $MI$ equal to 80%. The linearization model is based on the static regime SOA equations, and thus it has the advantage of lower complexity compared to a model based on dynamic SOA equations. This approach, however, is limited by the fact that the effectiveness of the method decreases as the frequency of the signal to be sampled increases. Nevertheless, a significant THD improvement is observed for sinusoidal signals to be sampled, at 250 MHz and 500 MHz for *MI* equal to 40%, 60% and 80%, and at 1 GHz for *MI* equal to 60%, while a tangible THD improvement is observed for the rest of frequency and *MI* combinations.

The quality of the sampling process depends on the signal to be sampled at the SOA-MZI input A, which acts as a reference signal and sets a hard limit on the maximum achievable performance by the SOA-MZI photonic sampler as well as on the efficiency of the post-distortion techique. Furthermore, adapting the SOAs design for specific use in an SOA-MZI photonic sampler so as to ensure stronger dependence of the SOA carrier density on the input power and higher saturation power for a wider range of acceptable input powers could also lead to performance improvements.

**Funding.** French state, Brittany region, European Regional Development Fund (CPER SOPHIE-Photonique-ATOM) and Brest Métropole (ARED Choraal).

**Acknowledgments.** We are grateful to Keysight for lending us an N1000A DCA-X sampling oscilloscope, without which the dynamic regime measurements would not have been possible.

**Disclosures.** The authors declare no conflicts of interest related to this article.

**Data availability.** Data underlying the results presented in this paper are not publicly available at this time but may be obtained from the authors upon reasonable request.